\documentclass[aps,reprint,nofootinbib,nobibnotes,notitlepage,superscriptaddress,onecolumn,prd,
 amsmath,amssymb
]{revtex4-2}

\usepackage[caption=false]{subfig}
\usepackage{braket}
\usepackage{float}
\usepackage{lipsum}
\usepackage{graphicx}
\usepackage{dcolumn}
\usepackage{bm}
\usepackage{natbib}
\usepackage{hyperref}
\hypersetup{
	colorlinks = true,
    linkcolor = Red,
    urlcolor  = Red,
    citecolor = Red
}
\usepackage{subfig}

\usepackage{microtype}

\usepackage{verbatim}
\usepackage[amssymb]{SIunits}
\usepackage{tabularx}
\usepackage[dvipsnames]{xcolor}
\usepackage{wasysym}
\usepackage[export]{adjustbox}
\usepackage{multirow}
\usepackage{tikz}

\def\be{\begin{equation}}
\def\ee{\end{equation}}
\usepackage{enumitem}

\usepackage{color}
\definecolor{darkgreen}{RGB}{0,120,0}

\definecolor{darkgreen}{RGB}{0,120,0}

\newcommand{\delD}[1]{(2\pi)^3\delta_\mathrm{D}\left({#1}\right)}

\newcommand{\av}[1]{\left\langle{#1}\right\rangle} 

\newcommand{\vk}{\vec k}
\newcommand{\hk}{\hat{\vec k}}

\newcommand{\F}{\mathcal{F}}

\newcommand{\hr}{\hat{\vec r}}

\newcommand{\fnl}{f_{\rm NL}}

\renewcommand{\vr}{\vec r}

\def\beq{\begin{eqnarray}}
\def\eeq{\end{eqnarray}}
\let\vec\mathbf

\usepackage{empheq}

\usepackage{booktabs}

\begin{document}

\title{{\LARGE Separating the Inseparable:\\}
{\large Constraining Arbitrary Primordial Bispectra with Cosmic Microwave Background Data}}

\author{Oliver~H.\,E.~Philcox}
\email{ohep2@cantab.ac.uk}
\affiliation{Leinweber Institute for Theoretical Physics at Stanford, 382 Via Pueblo, Stanford, CA 94305, USA}
\affiliation{Kavli Institute for Particle Astrophysics and Cosmology, 382 Via Pueblo, Stanford, CA 94305, USA}
\affiliation{Simons Society of Fellows, Simons Foundation, New York, NY 10010, USA}
\affiliation{Center for Theoretical Physics, Columbia University, New York, NY 10027, USA}
\author{Kunhao Zhong}
\affiliation{Department of Physics and Astronomy,
University of Pennsylvania, Philadelphia, PA 19104, USA}
\author{Salvatore Samuele Sirletti}
\affiliation{Kavli Institute for the Physics and Mathematics of the Universe (WPI), UTIAS, The University of Tokyo, Chiba 277-8583, Japan}
\affiliation{Dipartimento di Fisica e Scienze della Terra, Università degli Studi di Ferrara, via Saragat 1, I-44122 Ferrara, Italy}
\affiliation{Dipartimento di Fisica, Università di Trento, via Sommarive 14, Trento, 38123, , Italy}
\affiliation{Instituto Nazionale di Fisica Nucleare, Sezione di Ferrara, via Saragat 1, I-44122 Ferrara, Italy}

\begin{abstract} 
\noindent 
To efficiently probe primordial non-Gaussianity using Cosmic Microwave Background (CMB) data, we require theoretical predictions that are factorizable, \textit{i.e.}\ those whose kinematic dependence can be separated. This property does not hold for many models, hindering their application to data. In this work, we introduce a general framework for constructing separable approximations to primordial bispectra, enabling direct CMB constraints on arbitrary models including those computed using numerical tools. In contrast to other approaches such as modal decompositions, we learn the basis functions directly from the data, allowing high-fidelity representations with just a handful of terms. This is practically implemented using machine-learning techniques, utilizing neural network basis functions and a loss function designed to mimic the CMB cosine similarity. We validate our pipeline using a variety of input bispectra, demonstrating that the approximations are $>99.5\%$ correlated with the truth with just three terms. By incorporating the neural basis into the \textsc{PolySpec} code, we derive KSW-type CMB estimators, which reproduce local- and equilateral-type non-Gaussianity to within $0.1\sigma$. As a proof-of-concept, we constrain two inflationary bispectra from the `cosmological collider' scenario; these feature an additional strongly-mixed particle sector and cannot be computed analytically. By combining the numerical predictions from \textsc{CosmoFlow} with our factorizable approach (with just three terms), we place novel constraints on the collider models using \textit{Planck} PR4 data, finding no detection of non-Gaussianity. Our method facilitates detailed studies of the inflationary paradigm, connecting modern theoretical tools with high-resolution observational data.
\end{abstract}

\maketitle

\section{Introduction}\label{sec: intro}

\noindent What happened in the early Universe? Since we cannot observe the inflationary epoch directly \citep[e.g.,][]{Guth:1980zm,Starobinsky:1982ee}, this is a notoriously difficult question to answer. Our best hope is to look at inflationary remnants such as the metric perturbations at horizon exit, which survive today as large-scale density inhomogeneities and stochastic gravitational waves \citep[e.g.,][]{Weinberg:2003sw}. The tensor modes encode the characteristic energy scale of inflation, $E_{\rm inf}$; thus far, they have not been detected, leading to an upper bound of $E_{\rm inf}\lesssim 10^{15}\mathrm{GeV}$ \citep{Lyth:1996im,Tristram:2021tvh}. In contrast, inflationary scalar modes source curvature fluctuations, and have been carefully analyzed using data from the Cosmic Microwave Background (CMB) and Large Scale Structure (LSS) \citep[e.g.,][]{Planck:2018jri,Planck:2019kim,2006JCAP...05..004C,Philcox:2024wqx,Philcox4pt3,Sohn:2024xzd,Jung:2025nss,DAmico:2022gki,Cabass:2022wjy,Cabass:2022ymb,Cabass:2024wob,Chaussidon:2024qni}.

In the simplest models of inflation, involving a single field with negligible self-interactions, primordial perturbations obey Gaussian statistics, and can be fully described by their two-point functions \citep[e.g.,][]{Cabass:2016cgp,Maldacena:2002vr}. Many well-motivated models do not fall in this category, however, including those with large self-interactions, modified initial vacua, energy dissipation, and beyond \citep[e.g.,][]{Byrnes:2010em,Komatsu:2010hc,Senatore:2009gt,Salcedo:2024smn,Bartolo:2004if,Chen:2010xka,Babich:2004gb}. A particularly interesting possibility is the `cosmological collider' scenario, involving an additional (possibly massive and spinning) field, which leaves detectable imprints in the curvature distribution \citep{Arkani-Hamed:2015bza,Chen:2010xka,Lee:2016vti}. In general, non-linear effects such as the above generate higher-order correlation functions, implying that non-Gaussianity of the scalar curvature field (or tensor sector) is a sensitive probe of primordial physics. This observation has sparked a wide variety of studies, both predicting the shape and structure of inflationary three-point functions (and beyond) \citep[e.g.,][]{Arkani-Hamed:2015bza,Lee:2016vti,Senatore:2009gt,Senatore:2010wk,Arkani-Hamed:2018kmz,Chen:2017ryl,Babich:2004gb,Chen:2010xka}, and attempting to constrain them from observational CMB and LSS data \citep[e.g.,][]{Planck:2019kim,Planck:2015zfm,2014A&A...571A..24P,Marzouk:2022utf,Creminelli:2005hu,Philcox:2024wqx,Philcox4pt3,Philcox:2023xxk,Sohn:2024xzd,Sohn:2023fte,Cabass:2022wjy,Cabass:2022ymb,Cabass:2024wob,DAmico:2022gki,Chaussidon:2024qni}. To date, there have been no robust detections of primordial non-Gaussianity (PNG), which places strong bounds on a variety of inflationary models, both microphysical and phenomenological in origin.

To constrain a particular model of primordial physics, we require both an accurate computation of the associated $N$-point function (e.g., the bispectrum or three-point correlation function) and an efficient method for comparing data and theory. In the first case, the historical work-horse has been the in-in method (or Schwinger-Keldysch formalism), which allows tree-level correlators to be computed as integrals over inflationary time \citep[e.g.,][]{Maldacena:2002vr,Weinberg:2005vy,Chen:2017ryl}. More recently, a slew of alternative techniques have been introduced, including bootstrap methods, which derive the three- and four-point functions from symmetry principles \citep[e.g.,][]{Arkani-Hamed:2018kmz}, and numerical approaches, obtaining the correlation functions as the solution of differential equations \citep[e.g.,][]{Werth:2023pfl,Mulryne:2009kh,Dias:2016rjq}. These approaches side-step some of the practical difficulties involved with predicting bispectra and allow new regimes to be tested, such as strongly mixed (and non-perturbative) sectors.

Comparing such models to data can be a challenge. When performing CMB analyses, it is highly advantageous to have theoretical templates that are \textit{factorizable}, \textit{i.e.}\ those whose kinematic dependence can be separated. For the primordial bispectrum, $B_\zeta$ (which will be the focus of our work), this implies
\beq\label{eq: separable-bk}
    B_\zeta(k_1,k_2,k_3) \sim \sum_n \alpha_n(k_1)\beta_n(k_2)\gamma_n(k_3),
\eeq
for some scalar functions $\alpha_n,\beta_n,\gamma_n$. If the theoretical model of interest can be written in this form, it can be efficiently analyzed using KSW-type (Komatsu-Spergel-Wandelt) estimators, which scale as $\mathcal{O}(N_{\rm pix}\log N_{\rm pix})$ for $N_{\rm pix}$ CMB pixels \citep{Komatsu:2003iq}. If the model cannot be written in factorized form, the complexity of a direct bispectrum analysis increases to $\mathcal{O}(N_{\rm pix}^3)$. Though this can be partly ameliorated by performing binned bispectrum analyses \citep[e.g,][]{Bucher:2015ura}, integrating the theory model across such bins remains expensive.\footnote{Tensor non-Gaussianity provides an interesting counterexample, since the CMB signals are dominated by large-scales due to the transfer functions \citep{Shiraishi:2019yux,Philcox:2023xxk,Philcox:2024wqx}. Even in this case, compression into bins can be lossy however.} 

Whilst a few physical bispectrum models do satisfy condition \eqref{eq: separable-bk} (such as those with additional massless particles and self-interactions), the vast majority do not. To analyze such models, we must compute a separable \textit{approximation} to the bispectrum, which can then be analyzed via standard approaches. The simplest approach is to project the bispectrum onto the standard `local', `equilateral', and `orthogonal' templates (which are polynomials in $k_i$); the latter two provide a complete basis for self-interactions (at leading order), and can partially capture a number of other regimes \citep[e.g.,][]{Senatore:2009gt,Arkani-Hamed:2015bza}. Many models cannot be well-described by these templates however, such as: (1) massive collider shapes, which feature oscillations in the squeezed limit controlled by mass parameter $\mu$ \citep{Arkani-Hamed:2015bza,Meerburg:2016zdz,Lee:2016vti}:\footnote{These templates are intended only for demonstrative purposes, and do not include higher-order terms, normalizations, phases, \textit{et cetera}.}
\beq\label{eq: Bk-cosmo-coll}
    \lim_{k_1\ll k_3}\frac{B_\zeta(k_1,k_2,k_3)}{P_\zeta(k_1)P_\zeta(k_3)} \supset\left(\frac{k_1}{k_3}\right)^{3/2}\cos\left(\mu\log k_1/k_3+\varphi\right);
\eeq
(2) the low-speed collider, with an enhancement on intermediate scales depending on sound-speed parameter $\alpha$ \citep{Jazayeri:2023xcj}:
\beq
    B_\zeta(k_1,k_2,k_3) \supset\frac{P_\zeta(k_2)P_\zeta(k_3)}{1+(\alpha k_1^2/k_2k_3)^2}+\text{2 perms.};
\eeq
(3) resonant non-Gaussianity models, with equilateral oscillations induced by axion-like potentials with frequency $\omega$ \citep{Flauger:2010ja,DuasoPueyo:2023kyh}:
\beq
    B_\zeta(k_1,k_2,k_3) \supset\left(P_\zeta(k_1)P_\zeta(k_2)P_\zeta(k_3)\right)^{2/3}\sin\left(\omega\log (k_1+k_2+k_3)+\varphi\right);
\eeq
(4) folded non-Gaussianity from excited initial states at some initial time $\eta_0$ \citep{Holman:2007na,Meerburg:2009ys}:
\beq
    B_\zeta(k_1,k_2,k_3) \supset\left(P_\zeta(k_1)P_\zeta(k_2)P_\zeta(k_3)\right)^{1/3}\frac{\cos\left(\eta_0(k_1+k_2-k_3)+\varphi\right)}{k_1+k_2-k_3}+\text{2 perms.}
\eeq
For such templates, an alternative decomposition is clearly needed.

One way to proceed is to adopt the `modal' scheme \citep{2009PhRvD..80d3510F,2012JCAP...12..032F,Fergusson:2010gn,Sohn:2023fte,Regan:2010cn}, which decomposes a given target bispectrum into a polynomial basis (or a linear transformation thereof, e.g., a set of Legendre polynomials). Essentially, this introduces the general separable ansatz
\beq\label{eq: separable-bk-modal}
    B_\zeta(k_1,k_2,k_3) \sim \sum_{pqr} \alpha_{pqr}k_1^pk_2^qk_3^r,
\eeq
for integer $p,q,r$ (in the $n_s=1$ limit) and coefficients $\alpha_{pqr}$, which can be obtained using linear algebra. This was adopted by the \textit{Planck} collaboration to compute constraints on a number of simple non-Gaussianity templates (including the local-, equilateral- and orthogonal-templates) \citep[e.g.,][]{Planck:2019kim,Planck:2015zfm,2014A&A...571A..24P}, and a variant was recently used in the \textsc{cmb-best} framework to analyze a variety of cosmological-collider models (cf.\,\ref{eq: Bk-cosmo-coll}) \citep{Sohn:2023fte,Sohn:2024xzd}. Whilst this is a powerful approach, the resulting decompositions can be very high-dimensional (with \citep{Sohn:2024xzd} using $N\approx 5000$ terms in their collider analysis, and \citep{DuasoPueyo:2023kyh} omitting high-frequency resonant bispectra due to computational limitations), which makes efficient CMB analyses expensive. This occurs since the polynomial basis can be inefficient for many well-motivated templates, such as those presented above (though see \citep{Sohn:2023fte} for the use of oscillatory templates with known frequency).

How can we obtain a more efficient decomposition for a given bispectrum? In this work, we tackle the problem using basic machine learning techniques. In particular, we allow $\alpha_n(k),\beta_n(k),\gamma_n(k)$ to be free functions, which can be learned from the data. As we show below, this allows for \textit{considerably} lower-dimensional representations, with $N\lesssim3$ capturing the target bispectrum features at high fidelity. With an appropriate choice of loss function, this decomposition can be used to construct efficient CMB estimators for arbitrary bispectra, allowing for direct measurement of the underlying $\fnl$ amplitudes in minimal time. Importantly, this does not require an analytic form for the theoretical bispectrum, which allows us to constrain numerically computed templates, including the `cosmological collider' forms, outside the known analytic regimes (used in \citep{Sohn:2024xzd,Philcox4pt3}). Here, we demonstrate the approach on strongly-mixed models, that cannot be analyzed using standard methods. As a further extension, our decompositions could be used to run structure-formation simulations (building on the modal approach used in \citep{Anbajagane:2025uro,Anbajagane:2025xlt,2012PhRvD..86f3511F}).

In the remainder this paper, we introduce our bispectrum decomposition framework and present a proof-of-concept analysis deriving CMB constraints on numerically computed bispectrum templates (obtained using the \textsc{CosmoFlow} code \citep{Werth:2024aui}). In \S\ref{sec: theory}, we discuss the bispectrum model in more detail, before outlining the precise neural network architecture in \S\ref{sec: neural-networks}. \S\ref{sec: app} applies our code to several test cases (including factorizable shapes, for validation), before we demonstrate CMB constraints obtained using our decompositions in \S\ref{sec: cmb-constraints}. We conclude in \S\ref{sec: conclusion} with a discussion of future applications. Our bispectrum decomposition pipeline (including tutorials and an interface with \textsc{PolySpec}) is publicly available at \href{https://github.com/KunhaoZhong/separable_bk}{GitHub.com/KunhaoZhong/separable\_bk}.

\section{Separable Bispectrum Decompositions}\label{sec: theory}
\subsection{Primordial Correlators}
\noindent Given the Fourier-space primordial curvature perturbation $\zeta(\vk)$, we define the primordial power spectrum, $P_\zeta$, and bispectrum, $B_\zeta$, as
\beq\label{eq: PkBk-def}
    \av{\zeta(\vk_1)\zeta(\vk_2)} = \delD{\vk_1+\vk_2}P_\zeta(\vk_1), \qquad\av{\zeta(\vk_1)\zeta(\vk_2)\zeta(\vk_3)} = \delD{\vk_1+\vk_2+\vk_3}B_\zeta(\vk_1,\vk_2,\vk_3).
\eeq
Empirically, the power spectrum is well described by
\beq
    P_\zeta(k) = \frac{A_\zeta}{k^3}\left(\frac{k}{k_{\rm pivot}}\right)^{n_s-1}
\eeq
where $A_\zeta \equiv 2\pi^2 A_s \approx 4.1\times 10^{-8}$ is the dimensionless power spectrum amplitude, $n_s\approx 0.96$ is the slope (related to the slow-roll parameters), and $k_{\rm pivot} = 0.05\,\mathrm{Mpc}^{-1}$ is an (arbitrary) reference scale \citep{2020A&A...641A...6P}. For $n_s=1$, the power spectrum is scale-invariant; this corresponds to de Sitter space, which is assumed in most inflationary calculations. 

In most models of inflation, the primordial bispectrum satisfies certain properties, namely:
\begin{enumerate}
    \item \textbf{Reality}: $B_\zeta^*(-\vk_1,-\vk_2,-\vk_3) = B_\zeta(\vk_1,\vk_2,\vk_3)$. This holds since the curvature is a real field.
    \item \textbf{Parity}: $B_\zeta(-\vk_1,-\vk_2,-\vk_3) = B_\zeta(\vk_1,\vk_2,\vk_3)$. This holds since the curvature is a scalar.
    \item \textbf{Permutation Symmetry}: $B_\zeta(k_1,k_2,k_3)$ is invariant under any permutations of $\{k_1,k_2,k_3\}$ (which follows from definition \ref{eq: PkBk-def}). 
    \item \textbf{Isotropy}: $B_\zeta(\vk_1,\vk_2,\vk_3) = B_\zeta(k_1,k_2,k_3)$ for $k_i\equiv|\vk_i|$. This can be broken if the model contains a preferred direction (as in some gauge field models with non-vanishing backgrounds) \citep[e.g.,][]{Shiraishi:2013vja}. 
    \item \textbf{Scale Invariance}: If the inflationary physics is invariant under time-translations, the bispectrum satisfies
    \beq
    B_\zeta(\lambda k_1,\lambda k_2, \lambda k_3) =\lambda^{2(n_s-4)}B_\zeta(k_1,k_2,k_3),
    \eeq
    for scalar $\lambda$, such that $B_\zeta(k,k,k)\sim k^{-6}$ in the de Sitter limit. This can be broken in models with time-dependent couplings and features \citep[e.g.,][]{Hannestad:2009yx}.
\end{enumerate}
In the below, we will assume conditions (1)--(4), remaining agnostic to condition (5).

The curvature bispectrum is usually decomposed into a shape function, $S$, and an amplitude, $f_{\rm NL}$, motivated by the approximate scale-invariance described above. Explicitly, we define
\beq
    B_\zeta(k_1,k_2,k_3) \equiv \frac{18}{5}\fnl\left[P_\zeta(k_1)P_\zeta(k_2)P_\zeta(k_3)\right]^{2/3}S(k_1,k_2,k_3).
\eeq
where $S(k_1,k_2,k_3)$ is a symmetric function that satisfies $S(\lambda k_1,\lambda k_2,\lambda k_3) = S(k_1,k_2,k_3)$ for any $\lambda$, if property (5) holds above. 
By imposing $S(k_{\rm pivot},k_{\rm pivot},k_{\rm pivot})=1$, the non-linearity amplitude is defined as
\beq
    \fnl &\equiv& \frac{5}{18}\frac{k_{\rm pivot}^6}{A_\zeta^2}B_\zeta(k_{\rm pivot},k_{\rm pivot},k_{\rm pivot}).
\eeq
In the de Sitter limit ($n_s=1$), the above relations simplify to
\beq
    B_\zeta(k_1,k_2,k_3) \to \frac{18}{5}\fnl\frac{A_\zeta^2}{k_1^2k_2^2k_3^2}S(k_1,k_2,k_3), \qquad \fnl &\to& \frac{5}{18}\frac{k^6}{A_\zeta^2}B_\zeta(k,k,k),
\eeq
(for any $k$), though we will keep the more general definitions in this work. Two canonical bispectrum shape functions are the local and equilateral templates, defined as
\beq\label{eq: S-loc-equil}
    S_{\rm loc}(k_1,k_2,k_3) &=& \frac{1}{3}\left[\left(\frac{P_\zeta(k_1)P_\zeta(k_2)}{P_\zeta^2(k_3)}\right)^{1/3}+\text{2 perms.}\right]
    \\\nonumber
    S_{\rm eq}(k_1,k_2,k_3) &=& \left[\left(\frac{P_\zeta(k_1)}{P_\zeta(k_2)}\right)^{1/3}+\text{5 perms.}\right]-3\,S_{\rm loc}(k_1,k_2,k_3)-2.
\eeq
These simplify to functions of $k_1^2/(k_2k_3), k_1/k_2, 1$ (plus permutations) in the $n_s=1$ limit.

\subsection{Separable Approximations}
\noindent In this work, we seek separable approximations to arbitrary primordial bispectra, $B_\zeta$. Given the corresponding shape function, $S$, we define an $N$-term factorizable ansatz as
\beq\label{eq: factorized-B}
    S^{(\rm full)}_{\rm approx}(k_1,k_2,k_3) \equiv \frac{1}{6}\sum_{n=1}^N w_n\left[\alpha_n(k_1)\beta_n(k_2)\gamma_n(k_3) + \text{5 perms.}\right];
\eeq
for real basis functions $\alpha_n,\beta_n,\gamma_n$ and weights $w_n$; by construction, this is real, parity-symmetric, permutation invariant, and isotropic. We additionally consider the simpler cyclic form with $\beta_n=\gamma_n$:
\beq\label{eq: factorized-B-cyc}
    S_{\rm approx}^{(\rm cyc)}(k_1,k_2,k_3) \equiv \frac{1}{3}\sum_{n=1}^N w_n\left[\alpha_n(k_1)\beta_n(k_2)\beta_n(k_3) + \text{2 perms.}\right].
\eeq
Notably, these are not guaranteed to be scale-invariant, which is difficult to enforce without placing strong restrictions on the basis functions.

The local template defined in \eqref{eq: S-loc-equil} is exactly described by the cyclic model with just one term, with $\alpha = \{P_\zeta^{-2/3}\}$, $\beta = \{P_\zeta^{1/3}\}$ and $w = \{1\}$. Similarly, the equilateral template requires the full model with $N=3$: $\alpha=\{P_\zeta^{1/3},1,P_\zeta^{1/3}\}$, $\beta=\{P_\zeta^{-1/3},1,P_\zeta^{1/3}\}$, $\gamma=\{1,1,P_\zeta^{-2/3}\}$ with $w=\{6,-3,-2\}$. A similar decomposition can be found for the orthogonal shape.

For more general bispectra, we must carefully choose the basis functions entering \eqref{eq: factorized-B}. A simple choice would be polynomials, \textit{i.e.}\ $\alpha_n(k),\beta_n(k),\gamma_n(k) = k^{p_n},k^{q_n},k^{r_n}$ \textit{et cetera}. This approach, which underlies the modal decompositions \citep{2009PhRvD..80d3510F} ensures exact scale-invariance by requiring $p_n+q_n+r_n=0$, though can require a large number of basis elements to describe realistic models of interest \citep[e.g.,][]{Sohn:2024xzd,DuasoPueyo:2023kyh}. In this work, we instead use a multi-layer perceptron (\textit{i.e.}\ neural network) model for each basis function (with architecture discussed in \S\ref{sec: neural-networks}), which allows for greater flexibility, and thus smaller $N$. As above, this does not ensure scale-invariance (which is a general difficulty of non-monomial bases) -- here, we will instead learn this condition from the input bispectrum datavector. 

Several alternative choices are of interest. First, we could set $\gamma_n(k) = [\alpha_n(k)\beta_n(k)]^{-1}$; this is true for the monomial models discussed above (and ensures scale-invariance therein), but cannot capture many of the interesting templates discussed in \S\ref{sec: intro}. Secondly, we could allow the basis functions to be complex, adding a conjugate term to \eqref{eq: factorized-B} (motivated by the collider models \eqref{eq: Bk-cosmo-coll}, which include a factor $(k_1/k_3)^{i\mu}$). Since any complex function can be expressed in terms of a pair of real functions, this is equivalent to our fiducial model. Finally, we could assert additional symmetry by setting $\alpha_n=\beta_n=\gamma_n$ (or the cyclic model). These simplify the structure of each term in \eqref{eq: factorized-B} but may require a larger number of terms in the approximation (\textit{i.e.}\ increased $N$).

\subsection{Inner Product}
\noindent To learn the basis functions and weights entering \eqref{eq: factorized-B}, we require some metric to evaluate the efficacy of a given bispectrum approximation. Usually, one uses the bispectrum \textit{cosine}, defined as
\beq\label{eq: cosine}
    \cos(B,B') = \frac{\av{B|B'}}{\sqrt{\av{B|B}\av{B'|B'}}},
\eeq
given some inner product $\av{B|B'}$ \citep[e.g.,][]{2009PhRvD..80d3510F}. If $\cos(B_{\rm true},B_{\rm approx})\approx 1$, then the bispectrum approximation is accurate. For an idealized three-dimensional survey in primordial-space, an appropriate inner product is given by
\beq
    \av{B|B'}_{\rm 3D} &=& \int\prod_{i=1}^3\left[\frac{d\vk_i}{(2\pi)^3}\right]\delD{\vk_1+\vk_2+\vk_3}\frac{B(k_1,k_2,k_3)B'(k_1,k_2,k_3)}{P_\zeta(k_1)P_\zeta(k_2)P_\zeta(k_3)};
\eeq
this is proportional to the Fisher matrix between two bispectra \citep{2009PhRvD..80d3510F}. In terms of the shape functions, this can be written (following \citep{2009PhRvD..80d3510F})
\beq\label{eq: shape-3D}
    \av{S|S'}_{\rm 3D} &=& \int_{|k_2-k_3|\leq k_1\leq k_2+k_3}dk_1dk_2dk_3\,k_1^{(n_s-1)/3}k_2^{(n_s-1)/3}k_3^{(n_s-1)/3}S(k_1,k_2,k_3)S'(k_1,k_2,k_3)\\\nonumber
    &\approx& \int k^2 dk\,\int_{|x-y|\leq 1\leq x+y}dx\,dy\,S(1,x,y)S'(1,x,y)
\eeq
where we have dropped a normalization factor and integrated over the angular dependence analytically. In the second line, we have assumed the scale-invariant limit, which recovers the standard form \citep[e.g.,][]{2009PhRvD..80d3510F}.

In this work, our goal is to compute accurate bispectrum decompositions that can be used in CMB studies. In this case, the most relevant inner product would be the Fisher matrix between the CMB bispectra generated from the primordial bispectra; in practice, this is expensive to compute. Instead, we adopt a modified version of \eqref{eq: shape-3D}:
\beq\label{eq: shape-2D}
    \av{S|S'}_{\rm 2D} &=& \int_{|k_2-k_3|\leq k_1\leq k_2+k_3}dk_1dk_2dk_3\,\frac{W(k_1)W(k_2)W(k_3)}{k_1+k_2+k_3}S(k_1,k_2,k_3)S'(k_1,k_2,k_3),
\eeq
where the factor $(k_1+k_2+k_3)^{-1}$ converts from three dimensions (primordial curvature) to two (the CMB), following \citep{2009PhRvD..80d3510F}. We further introduce weighting functions, $W(k)$, which limit the integral to $k$-modes that appreciably impact the CMB. A simple choice would be to include all modes with $k\in[\ell_{\rm min}/\chi_{\rm rec},\ell_{\rm max}/\chi_{\rm rec}]$, where $\ell_{\rm min}=2$, $\ell_{\rm max}\approx 2000$ (for \textit{Planck}) and $\chi_{\rm rec}\approx 14\,000\,\mathrm{Mpc}$ is the distance to the last-scattering surface. Here, we adopt a more nuanced Wiener filtering scheme (adding a factor of $k$ for dimensional consistency):
\beq
    W(k) &\equiv & k\,P_\zeta(k)/N_\zeta(k), \qquad 
    N_\zeta(k) = \left(4\pi \sum_{\ell=\ell_{\rm min}}^{\ell_{\rm max}} (2\ell+1) \sum_{XY\in\{T,E\}}\mathcal{T}^X_\ell(k)[C_\ell+N_\ell]^{-1,XY}\mathcal{T}^Y_\ell(k)\right)^{-1}.
\eeq
$N_\zeta(k)$ is the reconstruction noise on $\zeta(\vk)$ inferred from a set of CMB observations with transfer functions $\mathcal{T}_\ell^{X}(k)$, signal $C_\ell^{XY}$, and noise $N_\ell^{XY}$, where $X,Y\in\{T,E\}$ indicate temperature and polarization. Practically, this restricts us to modes with $k_{\rm min} = 2\times 10^{-5}\,\mathrm{Mpc}^{-1}, k_{\rm max}=2\times 10^{-1}\,\mathrm{Mpc}^{-1}$, accentuating regimes with higher signal-to-noise due to acoustic oscillations.

The above inner product ensures that our bispectrum decompositions are sufficiently accurate for the problem of interest. We note that this does \textit{not} require $S_{\rm approx}\approx S_{\rm true}$ everywhere; instead, we require that the templates are similar in regimes that dominate the total signal-to-noise. For example, the local shape given in \eqref{eq: S-loc-equil} can be almost perfectly represented by $S_{\rm approx}(k_1,k_2,k_3) = (P_\zeta(k_1)/P_\zeta(k_2))^{1/3}/6+\text{5 perms.}$. In this work our focus is on CMB studies; for other applications (such as three-dimensional decompositions used to build simulations \citep{Anbajagane:2025uro,Anbajagane:2025xlt}), alternative loss functions may be preferred.

\section{Model Architecture}\label{sec: neural-networks}
\noindent In the previous section, we outlined our factorizable ansatz for the primordial bispectrum model, as well as the inner product used as a similarity metric. Here, we discuss the machine learning architecture used to obtain the free weights and functions ($w_n,\alpha_n,\beta_n,\gamma_n$) from a given input bispectrum.

A neural network in its simplest form consists of two basic components: an affine transformation (a matrix multiplication plus a bias term) and a nonlinear activation function. The matrix and bias parameters are learnable, meaning they are updated using gradients of the loss function. By stacking multiple layers of affine transformations and activation functions, a neural network builds a hierarchy of intermediate representations and can realize highly expressive function families. The universal approximation theorem further guarantees that such networks can approximate any continuous function on a compact domain— even with non-continuous activation functions—provided sufficient data and model capacity~\cite{2017arXiv170902540L}. In this work, we parameterize each of the functions $(\alpha_n, \beta_n, \gamma_n)$ as an individual neural network. Each network is multiplied by its corresponding learnable coefficient and combined to produce the full approximated shape function, with permutations included as defined in \eqref{eq: factorized-B} and \eqref{eq: factorized-B-cyc}.

For each $(\alpha_n, \beta_n, \gamma_n)$, we adopt a simple multi-layer perceptron (MLP) architecture with an additional scalar weight $w_n$. Each MLP contains three affine layers with $\tanh$ activations, and we optionally add a logarithmic transformation to the networks, setting $f(k)\to e^{\log f(k)}$. The outputs are then combined using products and permutation sums, which are fed into the loss function. We modify this directly in the \texttt{forward} function of \texttt{nn.modules} in \texttt{PyTorch}, which becomes auto-differentiable for training. Because we can easily generate a large number of training samples, this minimal MLP design performs well in practice. We also experimented with deeper networks, residual connections, and convolutional layers, but observed no improvement. The baseline model contains roughly 10,000 learnable parameters—very small compared to modern networks for high-dimensional tasks—which makes hyperparameter tuning straightforward. For most training-related hyperparameters, we rely on the default \texttt{PyTorch} settings, as varying them did not materially affect performance. Instead, the primary choices to tune are architectural: the number of terms ($N_{\rm terms}$) and the symmetry structure. A comparison of these model choices is presented in Sec.~\ref{subsec: prim-results}. Finally, we note that the training is performed hierarchically; starting from $N=1$ we train the network until saturation, then increase to $N=2$ (reusing the previously optimized parameters as starting points), and gradually increase until a desired accuracy threshold is reached (such as a $99.9\%$ cosine). We have publicly released the code and an associated tutorial on GitHub.\footnote{\href{https://github.com/KunhaoZhong/separable_bk}{GitHub.com/KunhaoZhong/separable\_bk}}

\section{Application to Primordial Models}\label{sec: app}
\subsection{Models}\label{subsec: models}
\noindent Next, we test our separable approximation scheme by applying it to a set of fiducial primordial bispectra with varying degrees of complexity. We consider four models:
\begin{enumerate}
    \item \textbf{Local}: We use the $S_{\rm loc}$ shape given in \eqref{eq: S-loc-equil} (setting $n_s=1$). This has a known decomposition with $N=1$, with signal-to-noise dominated by squeezed triangle configurations.
    \item \textbf{Equilateral}: We use the $S_{\rm eq}$ shape given in \eqref{eq: S-loc-equil} (also with $n_s=1$). Again there is an exact result, this time with $N=3$ and dominated by equilateral configurations.
    \item \textbf{Collider-I}: We use the bispectrum shape generated in an effective field theory two-field model of inflation, involving the usual Goldstone mode ($\pi$, \textit{i.e.}\ the inflaton) and a massive scalar $\sigma$ \citep[e.g.,][]{Senatore:2010wk,Arkani-Hamed:2015bza,Lee:2016vti}. Here, we consider the double-exchange bispectrum induced by the $\rho\dot\pi\sigma$ mixing (with strength $\rho$) and the $\dot\pi\sigma^2$ cubic interaction, following \citep{Pinol:2023oux}. We set $m_\sigma = 3H/2$, $\rho = 2H$ (for Hubble scale $H$), and assuming unit sound-speed. This shape was chosen since (a) it has limited overlap with the local, equilateral, and orthogonal templates; (b) it features oscillations near the squeezed limit; (c) it is `strongly-mixed' (with $\rho/H\gtrsim 1$), and thus cannot be computed analytically.
    \item \textbf{Collider-II}: Similarly to Collider-I, we assume a two-field model involving a massive scalar field. Here, we consider the single-exchange bispectrum with $\rho\dot\pi\sigma$ and $(\partial_i\pi)^2\sigma$ vertices following \citep{Jazayeri:2023xcj}, setting $m_\sigma=0.1H$, $\rho = 2H$, and sound-speed $c_s=0.05$. This falls in the regime of the `low-speed' collider, and was chosen since it (a) has an atypical enhancement outside the squeezed limit; (b) cannot be computed with perturbative methods due to strong-mixing; and (c) is not dwarfed by the associated self-interaction contribution (itself well-described by the equilateral template) \citep[Fig.\,9]{Jazayeri:2023xcj}. For this shape, a rough analytic approximation is available \citep{Jazayeri:2023xcj}; however, it is not separable except via a Mellin-Barnes integral (which would require large $N$ to evaluate). 
\end{enumerate}
We stress that these models certainly do not span the full gamut of cosmological collider physics; here, they are used only as representative examples.

For the collider shapes, we evaluate the bispectrum in the de Sitter limit using the \textsc{CosmoFlow} code \citep{Werth:2024aui} (based on \citep{Werth:2023pfl,Pinol:2023oux}). Specifically, we compute bispectra across $6000$ points in $(k_2/k_1,k_3/k_1)$-space, which are then used to build a three-dimensional interpolator, invoking scale invariance. Since computing the numerical predictions for highly squeezed triangles becomes expensive \citep{Werth:2024aui}, we restrict to ratios with $k_i/k_j\geq 10^{-3}$, supplementing with the analytic predictions outside this regime (from \citep{Arkani-Hamed:2015bza}). We further rewrite the rate-limiting step of \textsc{CosmoFlow} in \textsc{cython}, which reduces the runtime by a factor of around ten. 

For each template described above, we build a three-dimensional dataset using a grid in $(k_1,k_2,k_3)$-space for $k_i\in[k_{\rm min}, k_{\rm max}]$ (where the weight functions $W(k)$ are non-zero cf.\,\ref{eq: shape-2D}), restricting to triangles that satisfy momentum conservation. We adopt a logarithmic spacing for $k\in[k_{\rm min},10k_{\rm min}]$ (using $10$ terms) and a linear spacing for $k\in[10k_{\rm min},k_{\rm max}]$ (using $150$ terms), which ensures efficient training without neglecting squeezed configurations. The full data-vectors contain around $300\,000$ elements. Even if the bispectra of interest are scale-invariant, we note that it is \textit{imperative} to include data-points from the whole $k$-domain of interest in the training set to ensure that the machine learning architecture learns the approximate scale-invariance.

\subsection{Results}\label{subsec: prim-results}

\noindent To assess the performance of our pipeline, we applied it to the four bispectrum templates described in the previous section, adopting two distinct parameterization schemes. In the first (denoted as `linear transform'), the neural network directly learns the basis functions $\alpha_n(k)$, $\beta_n(k)$, $\gamma_n(k)$ entering \eqref{eq: factorized-B} and \eqref{eq: factorized-B-cyc}. In the second (denoted as `log transform'), we instead apply a transformation to the basis functions as discussed above; this can improve convergence and numerical stability for certain templates, particularly those dominated by squeezed configurations.

\begin{figure}
    \centering
    \includegraphics[width=\linewidth]{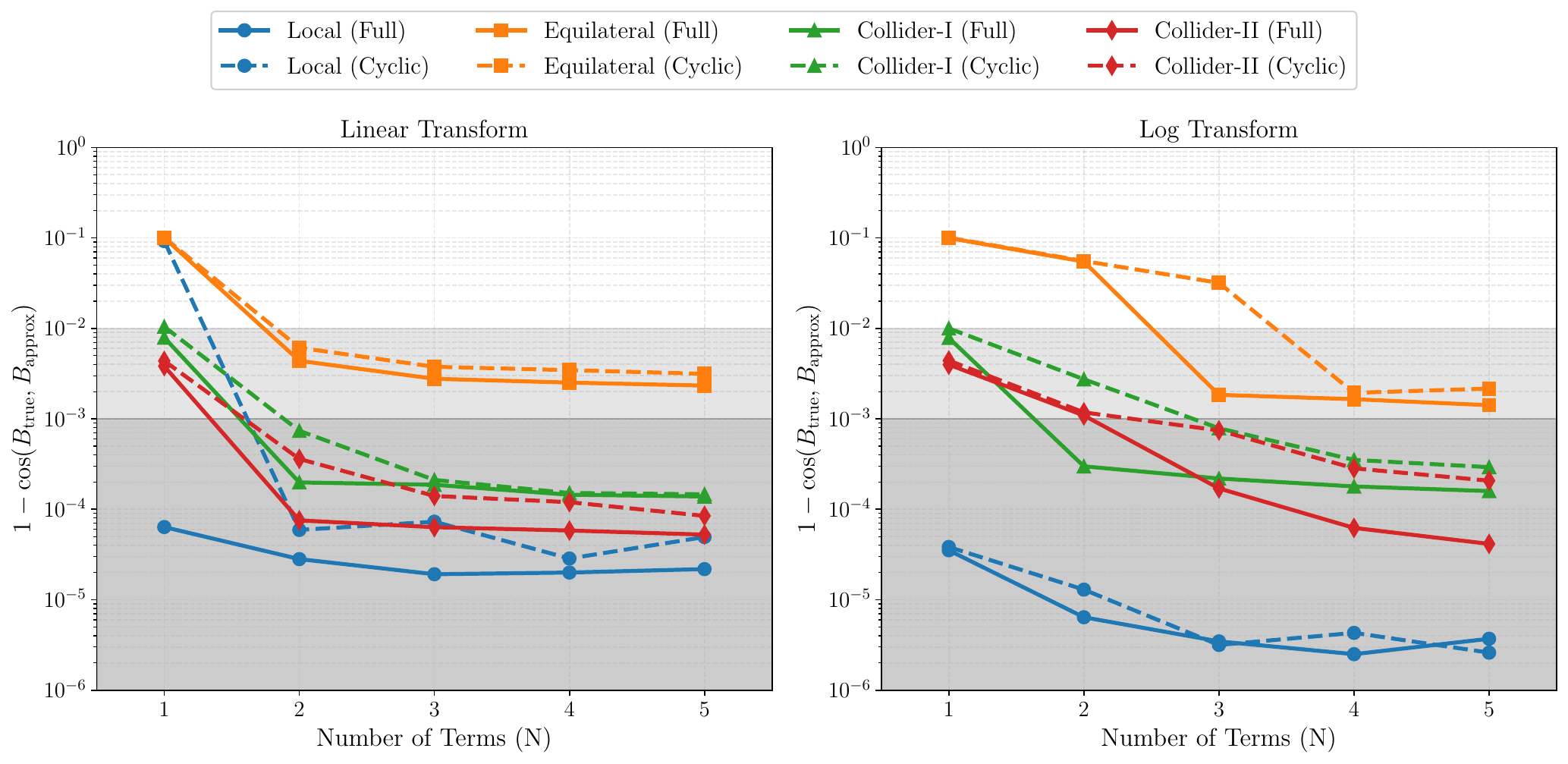}
    \caption{Convergence of the primordial-space bispectrum approximations as a function of the number of terms in the factorizable representation, $N$. We show results using both linear- (left) and log-transformed (right) neural networks for two types of permutation structure. For all templates considered (defined in \S\ref{sec: app}), we obtain high-accuracy approximations using $N\geq 4$, all with cosine similarities above $99\%$. Due to the differing template shapes, some decompositions are easier to learn than others, with, for example, the local shape being recovered at extremely high accuracy. The optimal choice of network architecture and symmetry assumptions also depends on the model in question, though all choices eventually lead to convergence. For the local and equilateral shapes, we find excellent results at $N=1$ (cyclic) and $N=3$ (full), recovering the standard analytic decompositions. For the collider shapes, which do not have an explicit analytic form, we can obtain $>99.9\%$ accuracy approximations at $N=3$, allowing for their efficient estimation from CMB data in \S\ref{sec: cmb-constraints}.}
    \label{fig:NN_results}
\end{figure}

For each template, we computed separable approximations with $N = 1, \ldots, 5$ terms, using both the full symmetry \eqref{eq: factorized-B} and cyclic symmetry \eqref{eq: factorized-B-cyc}. Rather than imposing a stopping criterion based on the cosine similarity value reaching a threshold (which is included by default in our code), we allowed the training to proceed through all values of $N$, enabling a systematic study of convergence behavior. Fig.~\ref{fig:NN_results} shows $1 - \cos(B_{\text{true}}, B_{\text{approx}})$ as a function of $N$ for all models and parameterization schemes. Across all cases, the pipeline demonstrates high efficiency, quickly reaching cosine values $>99\%$. Some templates are easier to learn than others; being strongly dominated by very few triangles (triangles with $k_{1,2}\sim k_{\rm max}$ and $k_3\sim k_{\rm min}$), the local template is the easiest, whilst the equilateral shape (dominated by $k_1\sim k_2\sim k_3\sim k_{\rm max}$) is the hardest. We also observe differing behavior for the two types of permutation structures and for linear or logarithmic transformations (although good results with all choices by $N\approx 4$). In practice, we suggest trying a variety of model assumptions to see which leads to the most accurate template at small $N$.

We note that the convergence curves are not strictly monotonic with increasing $N$, particularly for the local model. 
This arises due to the inherent randomness of the neural network training (via stochastic gradient descent). Moreover, we initialize with random weights, thus repeating the training procedure can yield slightly different convergence rates and final cosine values, though the approximations consistently converge to high fidelity. These sources of variability do not impact the practical utility of the method, as the achieved accuracies are more than sufficient for downstream CMB analyses.

In Tab.~\ref{tab:cosine_results}, we present the cosine similarities for representative values of $N$, chosen to match the expected dimensionality of each template (\textit{i.e.} $N = 1$ for local and $N = 3$ for equilateral shapes). 
For the local template, the logarithmic transform with cyclic symmetry achieves $\cos(B_{\text{true}}, B_{\text{approx}}) = 99.99\%$, consistent with the known analytic form \eqref{eq: S-loc-equil}. The equilateral template reaches cosines of $99.72\%$ ($99.82\%$) using a linear (logarithmic) transformation with the full symmetry structure, demonstrating accurate recovery of the expected three-term separable decomposition. For the cosmological collider models, which do not have exact analytic factorizations, the $N=3$ pipeline attains cosines exceeding $99.9\%$ across all parameterization schemes and symmetry types. This provides an excellent validation of our approach.

\begin{table}
\centering
\begin{tabular}{l|c|c|c|c}
\hline
\textbf{Template} & \textbf{$N$} & \textbf{Transform} & \textbf{Full Symmetry} & \textbf{Cyclic Symmetry} \\
\hline
Local & 1 & Linear & 0.9999 & 0.9077 \\
\cline{3-5}
      &   & Log    & 0.9999 & 0.9999 \\
\hline
Equilateral & 3 & Linear & 0.9972 & 0.9962 \\
\cline{3-5}
            &   & Log    & 0.9982 & 0.9681 \\
\hline
Collider-I & 3 & Linear & 0.9998 & 0.9998 \\
\cline{3-5}
           &   & Log    & 0.9998 & 0.9992 \\
\hline
Collider-II & 3 & Linear & 0.9999 & 0.9999 \\
\cline{3-5}
            &   & Log    & 0.9998 & 0.9993 \\
\hline
\end{tabular}
\caption{
Cosine similarity $\cos(B_{\rm true},B_{\rm approx})$ between true and approximated primordial bispectra. We show results for two classes of symmetry assumptions (\eqref{eq: factorized-B} and \eqref{eq: factorized-B-cyc}), optionally adding applying logarithmic transformations to the underlying neural basis functions. Results are shown for the Local, Equilateral, Collider-I, and Collider-II templates (as defined in \S\ref{sec: app}), and we fix the number of terms to $N=1$ (Local) and $N=3$ (Equilateral, Collider-I, and Collider-II). In all cases, we find excellent agreement between the true and approximated bispectra, validating our method.}
\label{tab:cosine_results}
\end{table}

\begin{figure}
    \centering
    \includegraphics[width=0.9\linewidth]{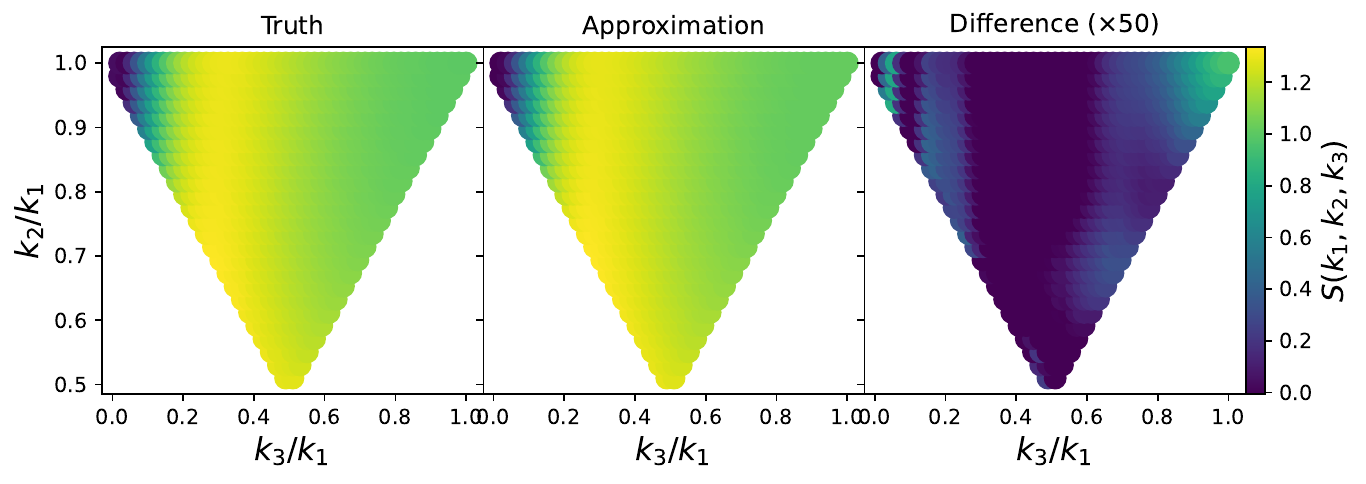}
    \includegraphics[width=0.9\linewidth]{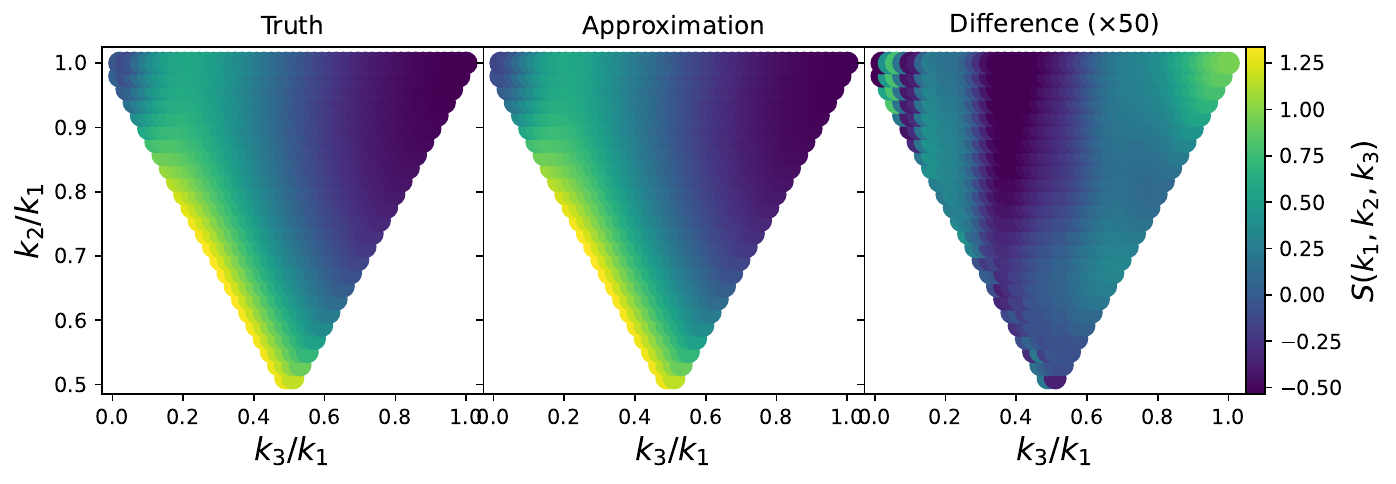}
    \caption{Comparison between the true (left) and approximated (middle) bispectrum shapes for the Collider-I model discussed in \S\ref{subsec: models}. The right panel shows the fractional difference, multiplying by $50$ for visibility. The factorizable approximation is calculated using $N=3$, assuming cyclic symmetries. The top panels show the primordial-space templates as a function of the dimensionless variables $k_2/k_1$ and $k_3/k_1$ (with $k_1 = 0.1\,\mathrm{Mpc}^{-1}$), with the top left, center bottom and top right edges corresponding to squeezed, flattened and equilateral configurations. In the bottom panel, we show the same results, but project out the equilateral template using \eqref{eq: deproj}. We find excellent agreement across the whole shape-space, with the factorizable representation achieving a cosine of 99.96\%.}
    \label{fig: primordial-coll}
\end{figure}

\begin{figure}
    \centering
    \includegraphics[width=0.9\linewidth]{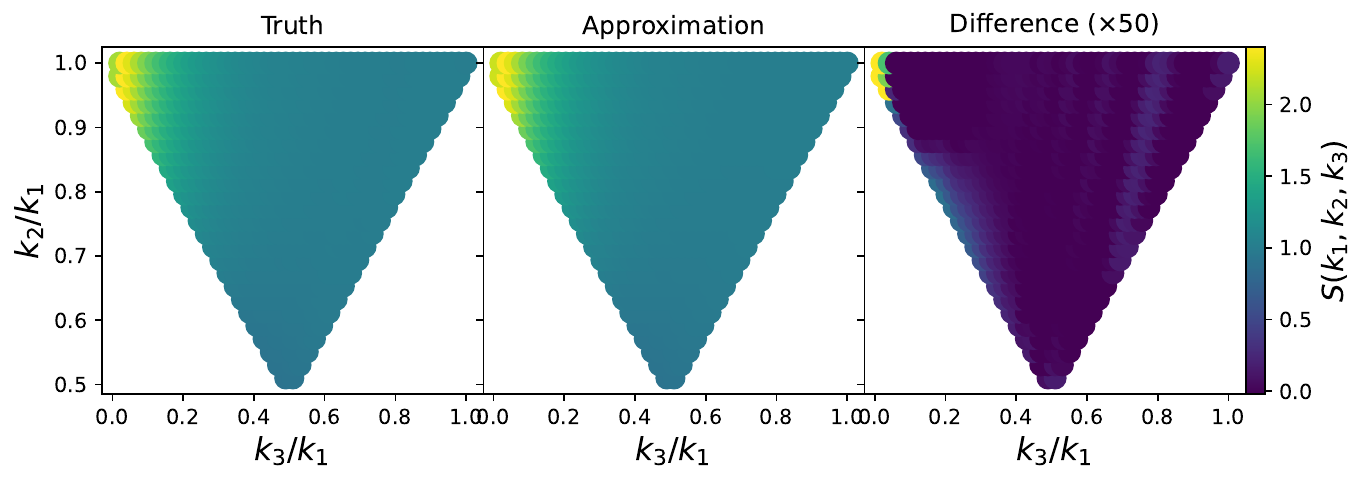}
    \includegraphics[width=0.9\linewidth]{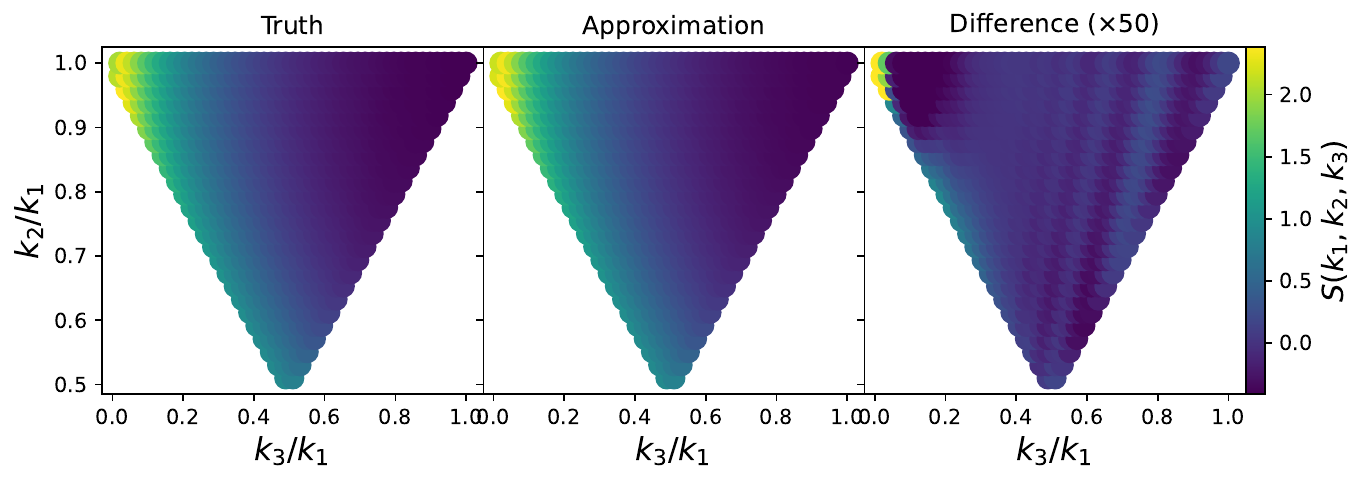}
    \caption{As Fig.\,\ref{fig: primordial-coll} but for the primordial Collider-II template. We again find excellent agreement, achieving a cosine of 99.95\%.}
    \label{fig: primordial-coll-slow}
\end{figure}

In the top panels of Figs.\,\ref{fig: primordial-coll}\,\&\,\ref{fig: primordial-coll-slow} we compare the true and approximated primordial shape functions obtained for the Collider-I and Collider-II models using our pipeline. For this purpose, we assume the cyclic template \eqref{eq: factorized-B-cyc} and use an optimization threshold of $99.9\%$, which requires $N=3$ for both templates (cf.\,Fig.~\ref{NN_results}). Across all of parameter space, including the equilateral ($k_1\approx k_2\approx k_3$), flattened ($k_1\approx k_2+k_3$) and squeezed ($k_3\ll k_1,k_2$) limits, we find excellent agreement between the true templates and the factorized forms, with differences visible at the $\sim 2\%$ level at most. Notably, both collider templates exhibit considerable ($\approx 80\%$) degeneracy with the equilateral shape,\footnote{This occurs since the equilateral shape encodes inflaton self-interactions, which can be mimicked by integrating-out a heavy field.} which is itself exactly factorizable. As such, a more thorough test of our pipeline can be obtained by projecting out the equilateral component of each template via
\beq\label{eq: deproj}
    S_{\rm deproj} = S-\frac{\langle S|S_{\rm eq}\rangle}{\langle S_{\rm eq}|S_{\rm eq}\rangle}S_{\rm eq}.
\eeq
In the bottom panels of Figs.\,\ref{fig: primordial-coll}\,\&\,\ref{fig: primordial-coll-slow}, we compare the deprojected templates, which once again show excellent agreement (with cosines of $99.8\%$ and $99.9\%$ for Collider-I and Collider-II respectively). This is a non-trivial test, since the deprojected spectra cannot be simply represented by polynomials, and constitute only part of the signal-to-noise of the training data. Together, these results provide an excellent validation of our pipeline.

\section{Application to CMB Estimators}\label{sec: cmb-constraints}
\noindent In the previous section, we have demonstrated that our factorization pipeline obtains accurate separable representations of complex bispectra with just a handful of terms. Thus far, however, we have considered only primordial bispectra, $B_\zeta$, which cannot be observed directly. In this section, we develop and apply practical CMB estimators for general bispectra, and use them to (a) validate our pipeline in practical observational settings and (b) constrain numerically-computed bispectrum models using real data.

\subsection{Theory}
\noindent Given our factorizable bispectrum ansatz \eqref{eq: factorized-B}, we can define efficient estimators for the corresponding bispectrum amplitude, $\fnl$. In general, the CMB bispectrum associated to a primordial signal $B_\zeta$ is given by \citep[e.g.,][]{Komatsu:2003iq,Komatsu:2010hc}
\beq\label{eq: cmb-bispec}
    \av{a^{X_1}_{\ell_1m_1}a^{X_2}_{\ell_2m_2}a^{X_3}_{\ell_3m_3}}_c = \prod_{i=1}^3\left[4\pi i^{\ell_i}\int\frac{d\vk_i}{(2\pi)^3}\mathcal{T}_{\ell_i}(k_i)Y^*_{\ell_im_i}(\hk_i)\right]\delD{\vk_1+\vk_2+\vk_3}B_\zeta(\vk_1,\vk_2,\vk_3),
\eeq
depending on the transfer functions $\mathcal{T}_\ell^X(k)$ and spherical harmonics $Y_{\ell m}(\hk)$, where $X_i\in\{T,E\}$ indexes temperature and polarization. By extremizing the likelihood for the CMB dataset, $a_{\ell m}^{X}$, with respect to $\fnl$, an optimal estimator is straightforwardly derived \citep{Komatsu:2003iq} (see also \citep{2011MNRAS.417....2S,Philcox4pt1}):
\beq\label{eq: opt-estimator}
    \widehat{f}_{\rm NL} &=& \frac{1}{6}\frac{1}{\mathcal{F}}\sum_{\ell_1\ell_2\ell_3m_1m_2m_3X_1X_2X_3}\frac{\partial\av{a^{X_1}_{\ell_1m_1}a^{X_2}_{\ell_2m_2}a^{X_3}_{\ell_3m_3}}_c}{\partial \fnl}\left(\tilde{a}_{\ell_1m_1}^{X_1}\tilde{a}_{\ell_1m_2}^{X_2}\tilde{a}_{\ell_3m_3}^{X_3}-3\tilde{a}_{\ell_1m_1}^{X_1}\av{\tilde{a}_{\ell_2m_2}^{X_2}\tilde{a}_{\ell_3m_3}^{X_3}}\right)^*
\eeq
where $\tilde{a}$ is an inverse-variance weighted CMB map, and $\mathcal{F}$ is a normalization factor, equal to the Fisher matrix. This projects three copies of the filtered CMB onto the theoretical model of interest, removing a linear term to suppress the large-scale variance (which is computed using simulations). \textit{A priori}, this is difficult to compute due to the coupled sum over $\ell_i,m_i$, which has $\mathcal{O}(\ell_{\rm max}^6)$ complexity. 

If the underlying bispectrum is separable, the estimator can be computed in $\mathcal{O}(\ell_{\rm max}^2\log\ell_{\rm max})$ time. Inserting the separable ansatz \eqref{eq: factorized-B} (with unit $\fnl$) and using \eqref{eq: cmb-bispec}, we can rewrite \eqref{eq: opt-estimator} as
\beq\label{eq: opt-estimator-factorized}
    \widehat{f}_{\rm NL} &=& \frac{3}{5}\frac{1}{\mathcal{F}}\sum_{n=1}^N w_n\int r^2dr d\hr\,\bigg[U[\tilde{a},\alpha_n](\vr)U[\tilde{a},\beta_n](\vr)U[\tilde{a},\gamma_n](\vr)\\\nonumber
    &&\qquad\qquad\qquad\qquad\qquad-\left(U[\tilde{a},\alpha_n](\vr)\av{U[\tilde{a},\beta_n](\vr)U[\tilde{a},\gamma_n](\vr)}+\text{2 perms.}\right)\bigg]
\eeq
defining
\beq
    U[a,f](\vr) &\equiv& \sum_{\ell mX}u_\ell^X[f](r)Y^*_{\ell m }(\hr)a_{\ell m}^{X*}, \qquad u_\ell^X[f](r) = \frac{2}{\pi}(-1)^{\ell}\int_0^{\infty}k^2dk\,f(k)P_\zeta^{2/3}(k)\mathcal{T}_{\ell}(k)j_{\ell}(kr)
\eeq
where $\vr\equiv(r,\hr)$. Since we have factorized the harmonic-space sums and $\vk$-integrals, this can be computed entirely using spherical harmonic transforms ($\sum_{\ell m}Y_{\ell m}$), one-dimensional integrals ($\int r^2dr$), and sums over pixels ($\int d\hr\,$). The canonical $\fnl^{\rm loc}$ estimator takes the above form \citep{Komatsu:2003iq}, with $u_\ell[\alpha_1]=u_\ell[\beta_1]=p_\ell$, $v_\ell[\gamma_1]=q_\ell$ in the notation of \citep{Philcox4pt1}. As discussed previously \citep{2011MNRAS.417....2S,Philcox4pt1,Philcox:2023uwe,Philcox:2023psd}, the normalization $\F$ can be computed as a Monte Carlo sum, involving the map derivative
\beq
    Q_{\ell m}^X[x,y] &=& \sum_{\ell_2\ell_3m_2m_3X_2X_3}\frac{\partial\av{a^{X}_{\ell m}a^{X_2}_{\ell_2m_2}a^{X_3}_{\ell_3m_3}}_c}{\partial \fnl}x^{X_2*}_{\ell_2m_2}y^{X_3*}_{\ell_3m_3}\\\nonumber
    &=& \frac{3}{5}\sum_{n=1}^{N}w_n\int r^2drd\hr\,u_{\ell}^{X}[\alpha_n](r)\int d\hr\,Y_{\ell m}^*(\hr)U[\beta_n,x](\vr)U[\gamma_n,y](\vr)+\text{5 perms.},
\eeq
which is computed using harmonic transforms, as before. To perform numerical optimization of the above estimators \citep[cf.][]{2011MNRAS.417....2S}, we require also the idealized Fisher matrix, defined by
\beq
    \mathcal{F}_{\rm ideal} &=& \frac{1}{6}\sum_{\ell_im_iX_iX'_i}\left(\frac{\partial\av{a^{X_1}_{\ell_1m_1}a^{X_2}_{\ell_2m_2}a^{X_3}_{\ell_3m_3}}_c}{\partial \fnl}\right)\left(\prod_{i=1}^3\tilde{C}_{\ell_i}^{-1,X_iX_i'}\right)\left(\frac{\partial\av{a^{X_1'}_{\ell_1m_1}a^{X_2'}_{\ell_2m_2}a^{X_3'}_{\ell_3m_3}}_c}{\partial \fnl}\right)^*,
\eeq
assuming unit mask and translation-invariant noise, where $\tilde{C}_\ell$ is the beam-deconvolved signal-plus-noise power spectrum \citep{2011MNRAS.417....2S}. Inserting the separable ansatz, we find
\beq
    \mathcal{F}_{\rm ideal} &=& (4\pi)^2\frac{9}{25}\sum_{nn'}w_nw_{n'}\int\frac{d\mu}{2}\int_0^{\infty}r^2dr\int_0^{\infty}r'^2dr'\\\nonumber
    &&\,\times\,\left(\zeta[\alpha_n,\alpha_{n'}](r,r',\mu)\zeta[\beta_n,\beta_{n'}](r,r',\mu)\zeta[\gamma_n,\gamma_{n'}](r,r',\mu)\,+\text{5 perms.}\right)
\eeq
defining
\beq
    \zeta[\alpha,\beta](r,r',\hr\cdot\hr') &=& \sum_{\ell m XX'} u_{\ell}^{X}[\alpha](r)\tilde{C}_{\ell}^{-1,XX'}u_\ell^{X'*}[\beta](r')Y^*_{\ell m}(\hr)Y_{\ell m}(\hr')\\\nonumber
    &\equiv& \sum_{\ell XX'} \frac{2\ell+1}{4\pi}u_{\ell}^{X}[\alpha](r)\tilde{C}_{\ell}^{-1,XX'}u_\ell^{X'*}[\beta](r')\mathcal{L}_\ell(\hr\cdot\hr'),
\eeq
for Legendre polynomial $\mathcal{L}_\ell(\mu)$. This can be efficiently computed using Gauss-Legendre quadrature.

\subsection{Code}
\noindent The general $\fnl$ estimator discussed above has been added to the public \textsc{PolySpec} package \citep{PolyBin} (see \citep{Philcox4pt2} for extensive discussion).\footnote{Available at \href{https://github.com/oliverphilcox/PolySpec}{GitHub.com/OliverPhilcox/PolySpec}.} This builds on the bispectrum and trispectrum estimators of \citep{Philcox4pt1,2011MNRAS.417....2S,Philcox:2023uwe,Philcox:2023psd,2015arXiv150200635S}, as well as the local-type and ISW-lensing bispectra discussed in \citep{Philcox:2025lxt}. Our implementation generalizes \eqref{eq: opt-estimator-factorized} to arbitrary linear weighting schemes (\textit{i.e.}\ choices of filtering in $\tilde{a}$), and corrects for the beam and mask via careful choice of the normalization matrix, $\F$. Importantly, the estimator is unbiased and, for suitable weighting, minimum variance (with $\mathrm{var}(\widehat{f}_{\rm NL})=\mathcal{F}^{-1}$) \citep{Philcox4pt1}. In practice, the code takes a set of input functions and weights ($\alpha_n,\beta_n,\gamma_n$ and $w_n$) obtained from the neural network model (\S\ref{sec: neural-networks}) and computes the relevant $U$ maps for each map and model of interest. Following \citep{2011MNRAS.417....2S,Philcox4pt1}, we pre-compute the $u_\ell(r)$ integrals from the input transfer function, using a dense array in $k$. We adopt a sparse (weighted) sampling in $r$ to compute the remaining integrals, itself determined via a numerical optimization procedure, using the derivatives of the ideal Fisher matrix $\F_{\rm ideal}$ with respect to $r,r'$ \citep{2011MNRAS.417....2S}.

\subsection{Practical Application}

\begin{figure}
    \centering
    \includegraphics[width=0.85\linewidth]{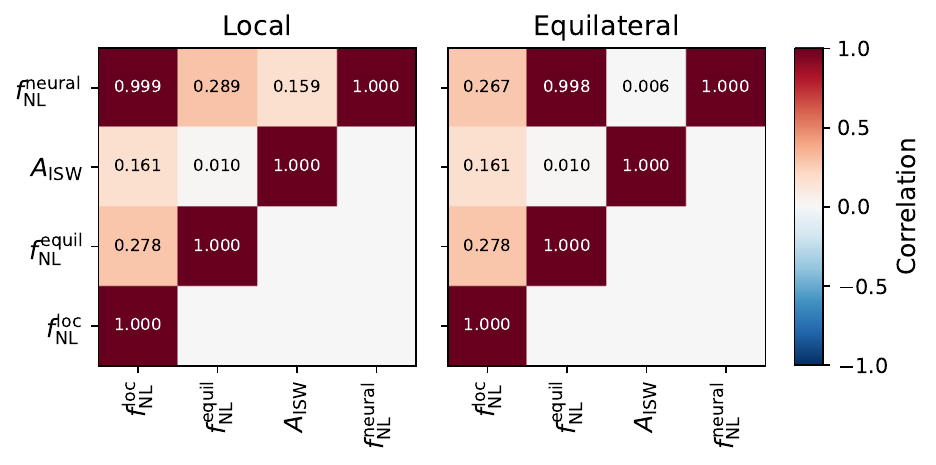}
    \caption{CMB correlation matrix for the neural, ISW-lensing, equilateral, and local $f_{\rm NL}$ amplitudes (where `neural' represents the factorized basis introduced in this work). The left (right) panel shows results using the factorized cyclic representation trained on the local (equilateral) shape with $N=1$ ($N=3$). Each value indicates the theoretical cosine between two templates in a \textit{Planck} analysis, including the observational mask, beam, and noise (computed using \textsc{PolySpec}). In both cases, the factorizable representations are almost perfectly correlated with the target shapes, and show similar correlations with the other shapes of interest. This is a non-trivial test of the pipeline since we compare two-dimensional CMB predictions, whilst the neural templates were trained only on three-dimensional correlators.}
    \label{fig: corner-fisher-cmb}
\end{figure}

\begin{figure}
    \centering
    \subfloat[Local Non-Gaussianity]{%
        \includegraphics[width=0.48\linewidth]{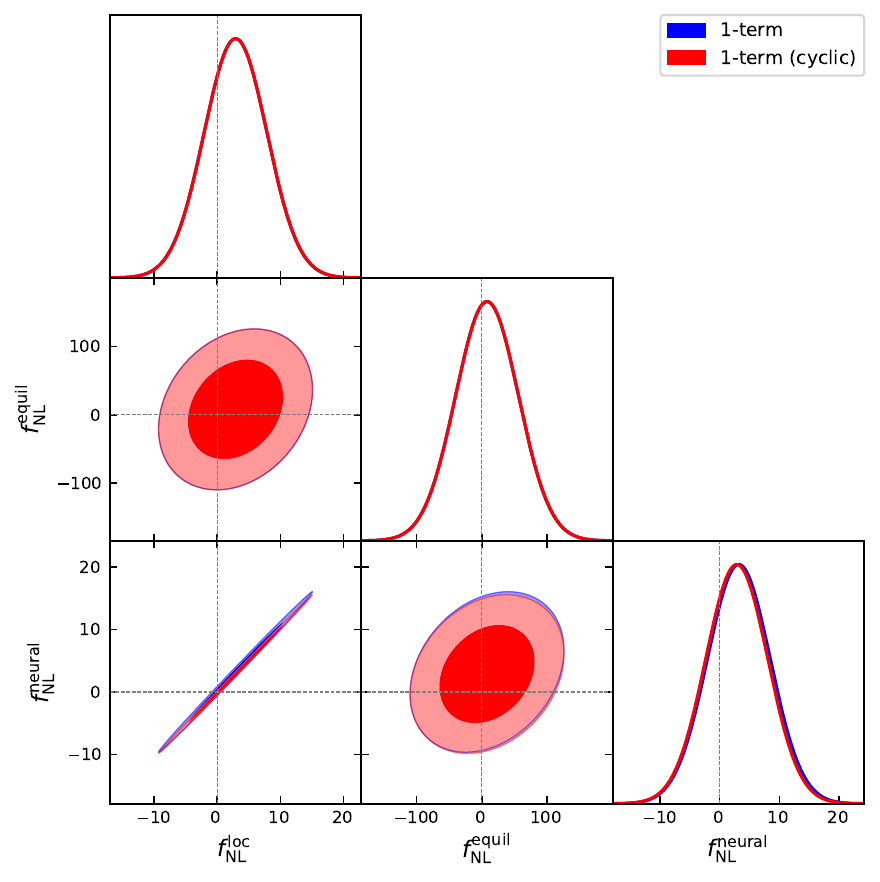}%
        \label{fig:corner_local2}%
    }%
    \hfill
    \subfloat[Equilateral Non-Gaussianity]{%
        \includegraphics[width=0.48\linewidth]{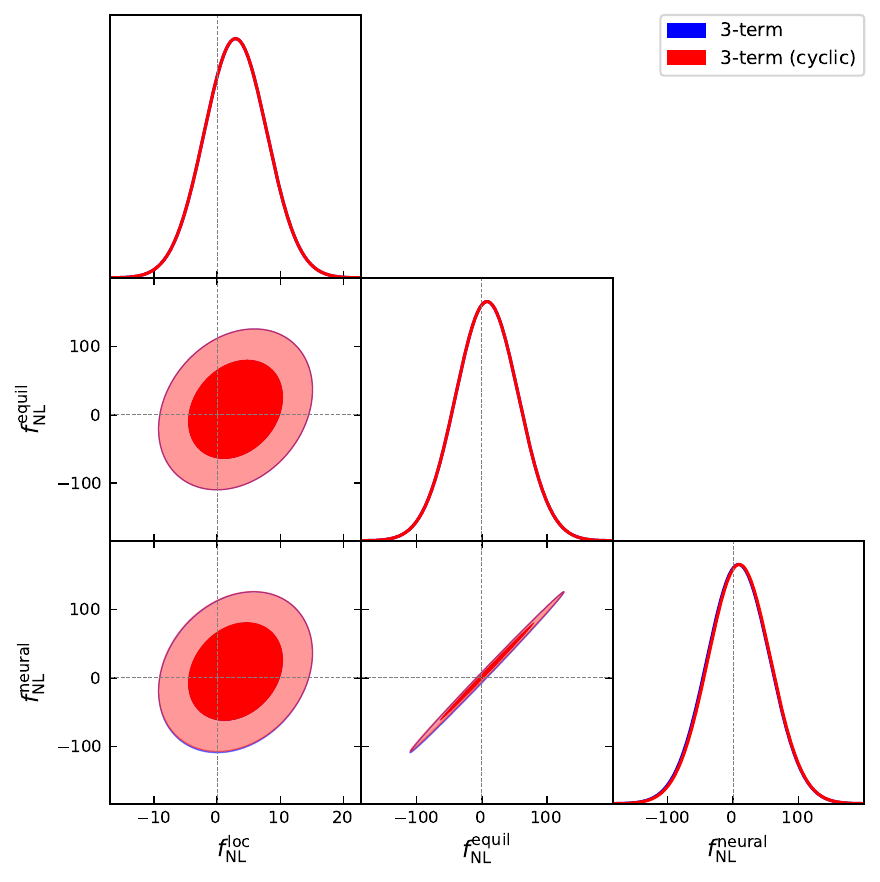}%
        \label{fig:corner_equil}%
    }
    \caption{Comparison of \textit{Planck} PR4 non-Gaussianity constraints obtained using standard templates (parametrized by $f_{\rm NL}^{\rm loc}, f_{\rm NL}^{\rm equil}$) to those obtained using the pipeline of this work (parametrized by $f_{\rm NL}^{\rm neural}$). In the left (right) panel, we use the machine-learning model to obtain a separable approximation to the local (equilateral) shape; if our pipeline is accurate, $f_{\rm NL}^{\rm neural}$ should exhibit strong degeneracies with $f_{\rm NL}^{\rm loc}$ ($f_{\rm NL}^{\rm equil}$). Here, each shape is analyzed independently, \textit{i.e.}\ the constraints are unmarginalized. We use $N=1$ for the local model and $N=3$ for the equilateral model, testing both full (blue) and cyclic (red) basis function expansions. Results are obtained using the full temperature-plus-polarization dataset, and errors are computed from 400 \textsc{FFP10}/\textsc{npipe} simulations. In all cases, we find excellent agreement between the standard constraints and those from our pipeline, validating our approach.}
    \label{fig: planck-loc-eq}
\end{figure}

\begin{figure}
\centering
    \subfloat[Collider-I]{%
        \includegraphics[width=0.48\linewidth]{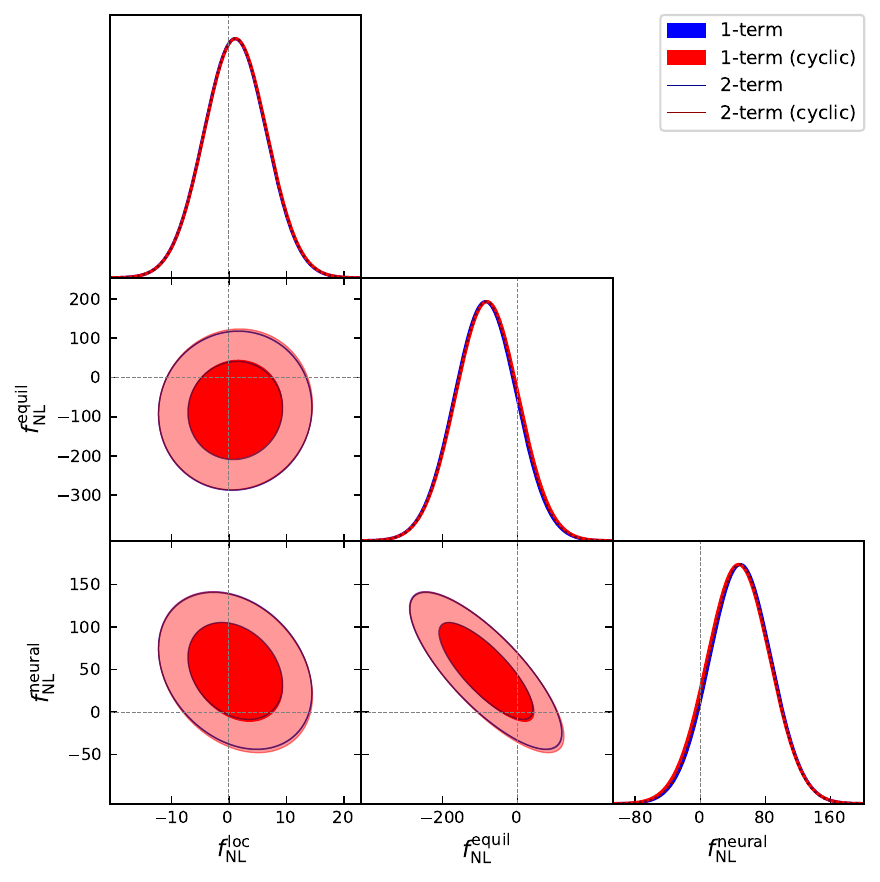}%
        \label{fig:corner_local}%
    }%
    \hfill
    \subfloat[Collider-II]{%
        \includegraphics[width=0.48\linewidth]{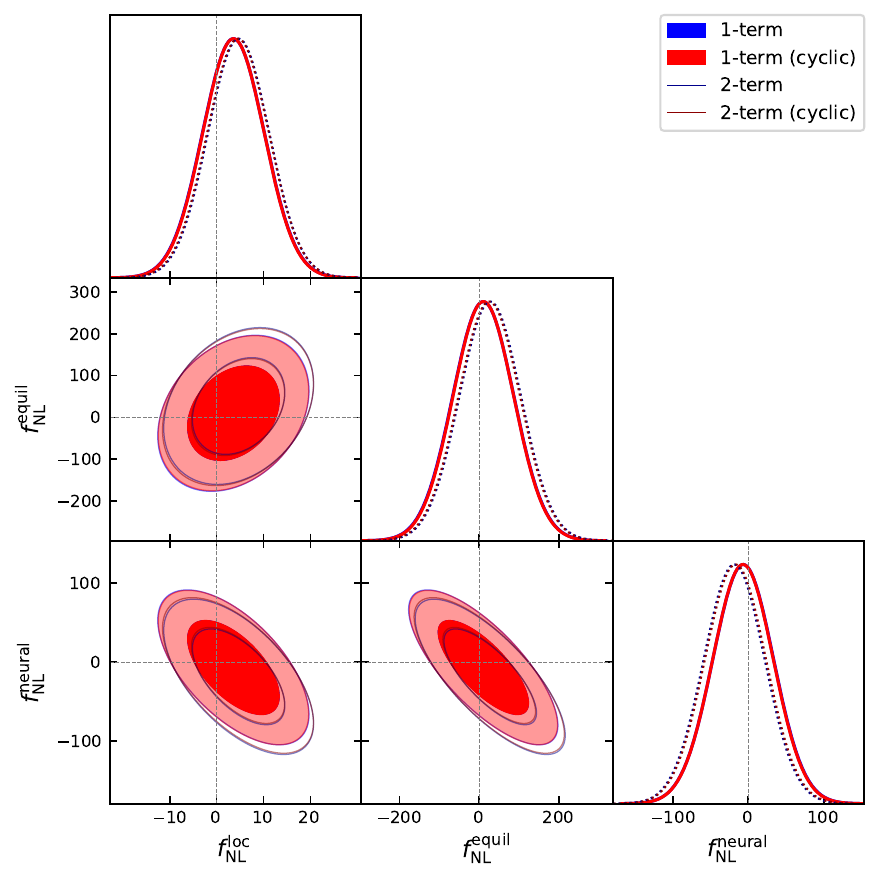}%
        \label{fig:corner_equil}%
    }
    \caption{\textit{Planck} PR4 constraints on the strongly-mixed cosmological collider models discussed in \S\ref{subsec: models}. Here, we perform a joint analysis of the collider templates with the local and equilateral shapes, additionally subtracting the expected ISW-lensing bias. All collider results are obtained using the separable template pipeline developed in this work, and we show results for both $N=1$ (solid) and $N=2$ (dashed) with full (blue) and cyclic (red) permutations. We find some correlations between the collider and equilateral templates in both cases (as expected) and find minimal dependence on the model hyperparameters. Utilizing the $N=2$ cyclic template, we find $f^{\rm neural}_{\rm NL}=48 \pm 38$ (Collider-I) and $f^{\rm neural}_{\rm NL}=-17\pm40$ (Collider-II), indicating no evidence for collider non-Gaussianity. These represent the first constraints on numerically-computed bispectra from CMB data.}
    \label{fig: planck-coll}
\end{figure}

\noindent To test our factorized estimators, we apply them to the public \textit{Planck} PR4/\textsc{npipe} dataset, as well as a suite of FFP10/\textsc{npipe} simulations \citep{Planck:2020olo,Planck:2015txa}. We adopt the \textsc{sevem} component-separation pipeline, and use $100$ simulations to form the linear term in \eqref{eq: opt-estimator}, with a further $400$ used to compute the estimator variance (similar to \citep{Philcox4pt3}). All data are masked with the \textit{Planck} common component-separated mask and filtered according to the total harmonic-space signal-plus-noise spectrum, with the latter obtained from half-mission maps, matching the official \textit{Planck} analyses \citep{Planck:2019kim}, and the PR4 re-analysis of \citep{Jung:2025nss}. In all cases, we include both temperature and polarization information. To avoid bias from lensing, we subtract off the expected ISW-lensing bias from all results \citep{Philcox:2025lxt,Planck:2019kim,Hill:2018ypf}.

We first perform a detailed analysis of the local and equilateral templates. As in \S\ref{sec: app}, we generate separable 3D decompositions for the local and equilateral templates using our pipeline, fixing $N=1$ and $N=3$ respectively. Utilizing the obtained basis functions and weights, we then compute the 2D bispectrum estimators using \textsc{PolySpec}, measuring the canonical $f_{\rm NL}^{\rm loc}$ and $f_{\rm NL}^{\rm equil}$ parameters alongside $f_{\rm NL}^{\rm neural}$, which encodes the amplitude of our factorizable representation, and the ISW-lensing amplitude $A_{\rm ISW}$. Using the estimator normalization matrix, $\F$ (which is the non-ideal Fisher matrix), we can compute the theoretical correlation matrix between the four templates, which shows the true CMB cosine between shapes, incorporating the observational mask, noise, and beam properties.

In Fig.\,\ref{fig: corner-fisher-cmb}, we show the correlations between the various shapes. When our separable decomposition is trained on the local shape (left panel), we find a 2D cosine of $99.93\%$ between $f_{\rm NL}^{\rm neural}$ and $f_{\rm NL}^{\rm loc}$; similarly, training on the equilateral shape (right panel) yields $99.81\%$. Moreover, we find that the neural representation accurately captures the correlations with other shapes with, for example, the local-ISW correlation equal to $16.10\%$ (true) or $15.94\%$ (neural approximation). These results imply that our decomposition is accurate, with the approximated CMB shape closely matching the true shape, even though the former was trained only in three-dimensional settings. 

In Fig.\,\ref{fig: planck-loc-eq}, we show the \textit{Planck} constraints on $f_{\rm NL}$ obtained from the local and equilateral shapes alongside those from our factorizable representations (analyzed independently). As expected we find almost perfect correlation between the true and approximated shapes, further noting that the amplitudes are highly consistent. For the setup considered herein, we obtain $f_{\rm NL}^{\rm loc}=2.9\pm5.0$ versus $f_{\rm NL}^{\rm neural}=2.9\pm5.2$ (left panel) or $f_{\rm NL}^{\rm equil}=8\pm48$ versus $f_{\rm NL}^{\rm neural}=9\pm48$ (right panel), again indicating that our approach is accurate. We additionally test the dependence on model set-up, finding that switching to the general permutation group \eqref{eq: factorized-B} instead of the cyclic model \eqref{eq: factorized-B-cyc} yields consistent results to within $0.1\sigma$.

Whilst the above results clearly demonstrate the validity of our approach, they are not particularly novel, given that the local and equilateral shapes are themselves already factorizable. To fully showcase the power of our approach, we must therefore apply the bispectrum decomposition and accompanying CMB estimators to (a priori) non-separable models. To this end, we perform a CMB analysis of the Collider-I and Collider-II models presented in \S\ref{subsec: models}, which do not have known analytic forms. To isolate the novel features of the collider templates, we perform a joint analysis of the local, equilateral, and collider models, subtracting ISW-lensing bias as before. To stress-test the model, we will use very low $N$, considering both $N=1$ and $N=2$ (motivated by Fig.\,\ref{NN_results}). 

Fig.\,\ref{fig: planck-coll} shows the corresponding constraints on strongly-mixed cosmological collider models for various choices of model hyperparameters. In all cases, we find no detection of new physics, with a maximum deviation of $1.3\sigma$. For the Collider-I (Collider-II) model, we find $f^{\rm neural}_{\rm NL}=48 \pm 38$ ($-17\pm40$), with $30\%$ and $80\%$ ($60\%$ and $75\%$) anticorrelation with the local and equilateral templates respectively. For the Collider-I shape, we find minimal dependence on the machine learning architecture with consistent results for $N=1$ and $N=2$ and both full and cyclic architectures to within $0.1\sigma$. For the Collider-II shape, we find a somewhat larger deviation, with a $0.3\sigma$ shift between the $N=1$ and $N=2$ models, though minimal dependence on the symmetry type. Such deviations could be reduced by using a more complex model, for example, setting $N=3$. Overall, we conclude that the factorizable basis can adequately represent complex numerical templates with just a few free basis functions, allowing such models to be analyzed in similar computation times to those of the standard local and equilateral templates.

\section{Discussion}\label{sec: conclusion}
\noindent In this work, we have proposed a novel pipeline to search for generic inflationary bispectra with CMB data. This works by decomposing a given input bispectrum shape onto a small set of factorizable basis functions, which can then be interfaced with standard KSW-type CMB estimators. Unlike conventional approaches (such as the modal scheme \citep{2009PhRvD..80d3510F,Sohn:2023fte}), our basis functions are learned from data itself rather than fixed. This yields a highly accurate model with just a handful of terms, in comparison to several thousand with standard approaches. To train this model, we have introduced a simple machine-learning pipeline, which trains free neural network functions via stochastic gradient descent, with a carefully chosen loss function designed to mimic the sensitivity of contemporary CMB experiments. Importantly, this allows us to directly compare any input inflationary model to observational data, leveraging recent developments in analytic and numerical inflation modeling.

We have rigorously tested our pipeline by applying it to a variety of bispectrum shapes. For the equilateral and local shapes (which have known separable forms), we find highly accurate results in primordial-space, with the true and approximate shapes obtaining cosines in excess of $99\%$. By inserting our decompositions into a CMB analysis pipeline (using the \textsc{PolySpec} code \citep{Philcox4pt2}) we have further demonstrated that our approach provides accurate and unbiased $f_{\rm NL}$ constraints, which are closely reproduce the standard $f^{\rm loc, equil}_{\rm NL}$ constraints (with cosine above $99.8\%$), despite the training only occurring in primordial space.

As a proof-of-concept, we have applied our techniques to two non-separable bispectrum models representing strongly-mixed cosmological colliders, optionally with small sound-speed \citep{Werth:2023pfl,Jazayeri:2023xcj,Arkani-Hamed:2015bza,Lee:2016vti}. These templates do not have known analytic forms and can only be computed numerically (here via \textsc{cosmoflow} \citep{Werth:2024aui}), which has thus-far impeded their practical analysis. Despite their interesting phenomenology (featuring oscillations both close-to and far from the squeezed limit), we have obtained factorizable representations of these shapes that are correlated with the full forms at $>99.5\%$ with just $\lesssim 3$ terms. Using this, we have placed the first observational constraints on the strongly-mixed cosmological collider using the \textit{Planck} PR4 data, finding no detections of $f_{\rm NL}$.

The cosmological collider scenarios considered herein represent only a tiny fraction of the possible inflationary models that can be probed with our tool. An important goal of future work will be to leverage such techniques to place detailed constraints on additional particles present during inflation, exploring novel regimes such as strong-mixing and low sound-speeds, going beyond the analytic lamppost (and thus previous works \citep{Sohn:2024xzd,Philcox4pt3}). Moreover, these techniques can facilitate analysis of other non-separable shapes, such as the dissipative models considered in \citep{Salcedo:2024smn}, the non-standard vacua scenarios of \citep{Meerburg:2009ys,Meerburg:2009fi}, or resonant effects, breaking scale-invariance \citep{Flauger:2010ja}. Furthermore, our basis could be extended to anisotropic bispectra or the tensor sector, for which efficient estimators are complex to derive \citep[e.g.,][]{Duivenvoorden:2019ses,Shiraishi:2019yux,Philcox:2024wqx}.

A number of additional extensions are possible. From the machine-learning side, it would be interesting to investigate whether alternative basis function parameterizations, such as those with Fourier features and complex components, could yield more efficient bispectrum decompositions. Moreover, one could try to build a common basis for a number of bispectrum shapes simultaneously; this would facilitate efficient parameter studies, scanning across physical quantities such as masses and sound-speeds. Furthermore, one could consider generalizing these approaches to the CMB trispectrum following the modal basis discussion in \citep{Regan:2010cn}. This statistic contains a wealth of information about primordial physics \citep[e.g.,][]{Philcox4pt1}, but has been largely unexplored, except in certain simplified template studies \citep{Philcox4pt3,Planck:2019kim,Marzouk:2022utf}.

Beyond the CMB, our decompositions could also be used in large-scale structure analyses. Whilst there is less need for efficient data analysis in this space (since one can use binned bispectra, which are fairly cheap to compute and analyze), one usually requires a separable bispectrum in order to generate the initial conditions for $N$-body simulations \citep{Scoccimarro:2011pz,Goldstein:2025eyj,Anbajagane:2025uro} (though see \citep{Fondi:2025xdf} for an alternative approach). By utilizing a similar decomposition to that discussed in this work (with a modified loss-function, perhaps down-weighting infrared contributions \citep[e.g.,][]{Scoccimarro:2011pz}), we could efficiently generate non-linear simulations with arbitrary models of inflationary non-Gaussianity, building upon the modal approaches considered previously \citep{Regan:2011zq,2012PhRvD..86f3511F}. As with the CMB applications, we expect that these methods will allow the inflationary paradigm to be probed at high sensitivity, particularly given the upcoming wealth of experimental data.

\vskip 8pt
\acknowledgments
{\small
\begingroup
\hypersetup{hidelinks}
\noindent 
We thank Thomas Bakx, Samuel Goldstein, and Sebastian Wagner-Carena for insightful discussions. OHEP was a Junior Fellow of the Simons Society of Fellows, and thanks Air Premia for hosting his visit. The computations in this work were run at facilities supported by the Scientific Computing Core at the Flatiron Institute, a division of the Simons Foundation. This paper was inspired by its title.
\endgroup
\vskip 4pt
}

\appendix

\bibliographystyle{apsrev4-1}
\bibliography{refs}

\end{document}